\documentclass[prl,showkeys,showpacs,twocolumn,preprintnumbers,amssymb,amsmath]{revtex4}
\usepackage{graphicx}
\usepackage{dcolumn}
\usepackage{bm}
\usepackage{latexsym,epsf,epsfig}
\usepackage{array}
\usepackage{pstricks}
\usepackage{caption}
\usepackage{amstext}

\begin{document}

\title{A Reappraisal of the Electric Dipole Moment Enhancement Factor for Thallium}
\author{H. S. Nataraj$^{1}$}\email{hsnataraj@gmail.com} \author{B. K. Sahoo$^2$} \author{B. P. Das$^3$ } \author{D. Mukherjee$^4$}
\affiliation{$^1$ Budapest University of Technology and Economics, 1111 Budapest, Hungary} 
\affiliation{$^2$ KVI, University of Groningen, NL-9747 AA Groningen, The Netherlands} 
\affiliation{$^3$ Indian Institute of Astrophysics, 560034 Bangalore, India}
\affiliation{$^4$ Indian Association for the Cultivation of Sciences, 700032 Kolkata, India}

\date{\today}
\begin{abstract}
The electric dipole moment (EDM) enhancement factor of atomic Tl is of considerable interest as it has been used in determining the most accurate intrinsic electron EDM limit to date. However, the value of this quantity varies from $-179$ to $-1041$ in different approximations. In view of the large uncertainties associated with many of these calculations, we have employed the relativistic coupled-cluster theory with single and double excitations and a subset of leading triple excitations and obtained the EDM enhancement factor of Tl as $-466$, which in combination with the most accurate measured value of Tl EDM yields $2.0 \times 10^{-27}\,{\mathrm e\,cm}$ as the new upper limit for the EDM of the electron. The importance of all-order correlation effects is emphasized and their trends are compared with those of two other ab initio calculations.
\end{abstract}


\maketitle

The EDM of a nondegenerate physical system \cite{ramsey50, landau57} arises from the simultaneous violations of parity ({$\mathcal P$}) and time-reversal (${\mathcal T}$) symmetries. The invariance of ${\mathcal T}$-symmetry is also associated with the invariance of the $\mathcal {CP}$ (combined charge conjugation (${\mathcal C}$) and ${\mathcal P}$ symmetries) symmetry on the basis of the $\mathcal {CPT}$ theorem \cite{schwinger}. Thus, EDMs of atoms can shed light on $\mathcal {CP}$ violations originating in the leptonic, semi-leptonic and hadronic sectors \cite{barr,pilaftsis}, some of which are not well understood so far. The knowledge of EDMs provide valuable insights into 
some profound questions such as the existence of new physics beyond the 
standard model (SM) and the matter-antimatter asymmetry in the universe \cite{barr,pilaftsis,kazarian}. Given the current interest in understanding different types of $\mathcal {CP}$ 
violations originating from elementary particles both by accelerator and 
non-accelerator based approaches, EDM searches are of great significance.

The EDMs of paramagnetic atoms are sensitive to the the EDM of the electron
\cite{sandars}. 
The most accurate limit for the latter has been obtained by combining the 
results of the Tl EDM measurement and the EDM enhancement factor (EF), defined as the ratio of the EDM of the atom to that of the 
electron, of this atom. Liu and Kelly have calculated this EF, and found its value to be $-585$, with an error bar of $5-10\%$, using a linearized version of the relativistic coupled-cluster (RCC) theory 
 \cite{liu92}. A more recent 
calculation of this quantity for Tl by Dzuba and Flambaum uses a hybrid approach combining the configuration interaction (CI) method and 
many-body perturbation theory (MBPT) and reports $-582$ with an estimated accuracy of 3\% 
\cite{dzuba2009}. Both these results, coincidentally, are in good agreement with each other. Nevertheless, considering the fact that the values of  
the Tl EDM EF reported in the literature range from $-179$ to 
$-1041$ \cite{hartley90,kraftmakher88,johnson86,flambaum76,sandars75}, there is clearly a need for high precision 
 calculations of this quantity. Thus, the primary focus of this 
paper is to determine the EF of Tl by proceeding beyond 
\cite{liu92,dzuba2009}. An accurate treatment of the unusually strong electron correlation 
effects in the ground state Tl EF warrants the use of an   
all-order relativistic many-body
method like the RCC theory.

The open-shell RCC theory with single, double and a subset of leading triple 
excitations employed in the calculation of the EDM EFs is 
discussed in detail in \cite{sahoo1,nataraj2,mukherjee}. However, we briefly present below the salient features of this method for the sake of completeness. The effective one electron form of the interaction Hamiltonian due to the electron EDM is given by \cite{mukherjee},
\begin{eqnarray}
H_{EDM}^{eff}\, = \, 2\, i\, c\, \sum_j \beta_j\, \gamma_j^5 \, \vec {p}_j\,^2
\label{eqn1} 
\end{eqnarray}
where $\beta$ and $\gamma^5$ are the usual Dirac matrices, $\vec p_j$ is the momentum 
vector of the $j^{th}$ electron. 
As the strength of the EDM interaction is 
sufficiently weak, we consider the wavefunction expansion only up to first-order in perturbation. Thus, the modified atomic wavefunction for a valence electron `$v$' is given by,
\begin{eqnarray}
|\Psi_v' \rangle = |\Psi_v^{(0)} \rangle + \left(\frac{d_e}{e\,a_0}\right)\, |\Psi_v^{(1)} \rangle\,.
\end{eqnarray}

In the RCC theory, the unperturbed and perturbed wavefunctions can be expressed as \cite{sahoo1,nataraj2,mukherjee},
\begin{eqnarray}
\vert \Psi_v^{(0)} \rangle &=& e^{T^{(0)}}\{1 + S_v^{(0)} \} \vert\Phi_v\rangle
\label{eqn3} \\
\text{and} \ \ \
\vert \Psi_v^{(1)} \rangle &=&  e^{T^{(0)}}\{ T^{(1)} \left (1 + S_v^{(0)} \right ) + S_v^{(1)} \} \vert\Phi_v\rangle \ \ \
\label{eqn4}
\end{eqnarray}
where $\vert\Phi_v\rangle$ is the Dirac-Fock (DF) wavefunction obtained by appending the valence 
electron $v$ to the closed-shell ($[5d^{10}]\,6s^2$) reference wavefunction, $T^{(0)}$ and $S_v^{(0)}$ are the excitation operators for core and valence electrons in an unperturbed case, where as, $T^{(1)}$ and $S_v^{(1)}$ are their first order corrections. The atomic wavefunctions are calculated using the Dirac-Coulomb Hamiltonian given by,
\begin{eqnarray}
H_0 &=& \sum_i \{ c \alpha_i \cdot p_i + (\beta_i - 1)m_i c^2 + V_{n}(r_i)\} + \sum_{i<j} V_C(r_{ij}), \nonumber \\ & & 
\label{H0}
\end{eqnarray}
where $\alpha$ and $\beta$ are Dirac matrices, $V_{n}(r_i)$ is the 
nuclear potential and $V_C(r_{ij})$ is the Coulomb potential on the electron $i$ due to the
$j^{th}$ electron.

We consider only the single and double excitation operators in the expansion of the RCC wavefunctions (termed as CCSD approximation), by defining, 
\begin{eqnarray}
T = T_1 + T_2 \hspace{0.5cm} \text{and} \hspace{0.5cm} S_v = S_{1v} + S_{2v},
\label{eqn5}
\end{eqnarray}
for both the perturbed and unperturbed operators. Further, we construct triple
excitation operators for $S_v^{(0)}$ as, 
\begin{eqnarray}\label{s30}
S_{vab}^{pqr,(0)} &=&  \frac{\widehat{H_0\, T_2^{(0)}} + \widehat{H_0\, S_{2v}^{(0)}}}{\epsilon_v + \epsilon_a + \epsilon_b - \epsilon_p - \epsilon_q - \epsilon_r},
\end{eqnarray}
 which are used to evaluate CCSD amplitudes iteratively. This is referred to as CCSD(T) method. Here, $\epsilon_i$ is the single particle energy of an orbital $i$. 

The final expression for the EDM EF (${\mathcal R} = \frac{D_a}{d_e}$) in 
terms of the coupled-cluster operators is given by,
\begin{eqnarray}\label{enfact}
{\mathcal R} &=& {\displaystyle \frac{1}{ea_0} \frac{\langle \Phi_v |\{1+ S_v^{(0)^\dagger}\} \overline{D^{(0)}} \{ T^{(1)} (1+ S_v^{(0)}) + S_v^{(1)}\} |\Phi_v \rangle }{\langle \Phi_v |\, e^{T^{(0)^\dagger}}\, e^{T^{(0)}} + S_v^{(0)^{\dagger}}\, e^{T^{(0)^{\dagger}}}\, e^{T^{(0)}}S_v^{(0)}\, | \Phi_v \rangle}}  \nonumber \\ &+& \text{h.c.}
\end{eqnarray}
where the dressed operator $\overline{D^{(0)}} = e^{{T^{(0)}}^\dagger}\, \vec{D}\, e^{T^{(0)}}$ and $\vec {D} = e\, \vec{r}~$ is the electric dipole moment 
operator due to the applied electric field. The procedure for the calculation of the 
above expression is discussed elsewhere \cite{sahoo1,nataraj2,mukherjee}.

\begin{table}[h]
  \caption{The contributions from various RCC terms to the EDM EF (${\mathcal R}$) of Tl. A new quantity, $X = T + S_v$ is defined to compare our results with those of \cite{liu92}.}\label{ef-tl-df-rcc}
 \begin{center}
\begin{tabular}{lcc}
 \hline \hline
Term &    This Work & Liu \& Kelly \cite{liu92} \\ \hline
$(D\, T_{1}^{(1)} )_{\text {lowest-order}}$   & $-153.6$   & $-153.2$ \\
$(D\, S_{1v}^{(1)} )_{\text {lowest-order}}$ & $-268.5$ & $-267.3$ \\ \hline
$(D\, T_{1}^{(1)})_{\text {higher-order}} $  &  $-224.7$ & $-342.1$ \\
$(D\, S_{1v}^{(1)})_{\text {higher-order}} $ & $-45.5$ & $-102.5$ \\ 
$D\, X_{2}^{(1)} $ & $248.0$ & $240.9$ \\ 
$X_{1}^{{(0)}^{\dagger}} D\, X_{1}^{(1)} $ & $22.5$ & $22.4$ \\ 
$X_{2}^{{(0)}^{\dagger}} D\, X_{1}^{(1)} $ & $-78.2$ & $49.3$ \\ 
$X_{2}^{{(0)}^{\dagger}} D\, X_{2}^{(1)} $ & $21.5$ & $-36.9$ \\ 
$X_{2}^{{(0)}^{\dagger}} D\, X_{1}^{(1)} X_{2}^{(0)} $ & $-4.3$ & $-2.2$ \\ 
\text{Higher order RCC terms} & $13.0$ & $-$ \\ 
\text{Normalization contribution} & $3.8$ & $6.5$ \\ \hline
Total EDM EF & $-466$ & $-585$ \\ \hline \hline
\end{tabular}
 \end{center}
\end{table}
\begin{figure}[h]
\includegraphics[width=6.5cm,clip=true]{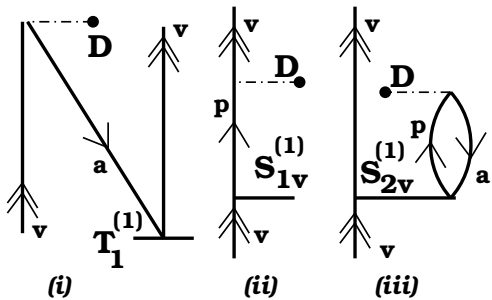}
\caption{The leading correlation diagrams: (i) ${D}\, T_1^{(1)}$, (ii) ${D}\, S_{1v}^{(1)}$, (iii) ${D}\, S_{2v}^{(1)}$. The exchange and hermitian conjugate diagrams are not shown. Labels $v$, $a$ and $p$ refer to valence, core and virtual orbitals, respectively.}
\label{fig1}
\end{figure}

In Table \ref{ef-tl-df-rcc}, we present the contributions from
different RCC terms along with the lowest order (DF) contributions to ${\mathcal R}$ and compare 
them with the results reported by Liu and Kelly \cite{liu92}.
It is evident from
this table that the bulk of the contributions to ${\mathcal R}$ comes from the RCC terms $D\, T_1^{(1)}$, $D\, S_{1v}^{(1)}$ and 
$D\, S_{2v}^{(1)}$. These important all-order correlation effects involving the core, valence and core-valence sectors are shown diagrammatically in Fig.\ref{fig1}. The single largest contribution ($-\,378$),  to ${\mathcal R}$ comes from the all-order core correlation effect of which its lowest order contribution is less than half of its magnitude. The all-order valence correlation contribution from $D\, S_{1v}^{(1)}$ is 
$-\,314$ while its DF contribution is $-\,269$. The difference between the all-order and the lowest-order
contributions of the terms $D\, S_{1v}^{(1)}$ and $D\, T_1^{(1)}$ clearly
indicate that the all-order core correlations are of crucial importance in the case of Tl. It is indeed significant  that one of the classes of the all-order core-polarization effects represented by $D\, S_{2v}^{(1)}$ makes the largest positive contribution, of magnitude $248$. This dramatically reduces the 
final result. There is a non-negligible contribution from the other higher order terms; however, many of them cancel each other to give an effective value of only about 13. The normalization of the RCC wavefunction also gives a contribution of 4 and the total result for the ground state EDM enhancement factor for Tl due to the intrinsic EDM of the electron amounts to $-466$.

\begin{figure}[h]
\includegraphics[scale=0.6,clip=true]{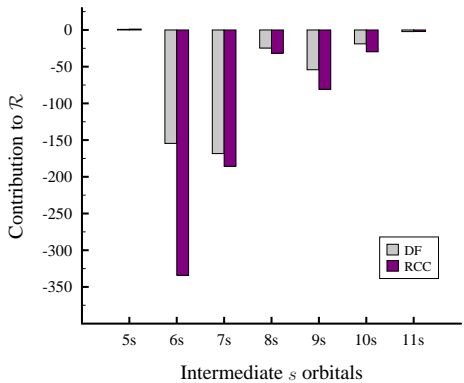}
\caption{The contributions from singly excited intermediate states, $5s$ through $11s$, in DF and RCC approximations are compared.}
\label{fig2}
\end{figure}

The $H_{EDM}^{eff}$ being an odd-parity operator mixes the atomic states of opposite parities, however, with the same angular momentum. Therefore, we have investigated the role of various intermediate states of $s$ symmetry. The RCC terms such as: $D\, T_1^{(1)}$, $D\, S_{1v}^{(1)}$, $S_{1v}^{{(0)}^{\dagger}} D\, S_{1v}^{(1)}$ and $S_{2v}^{{(0)}^{\dagger}} D\, S_{1v}^{(1)}$, have got significant contributions from the orbitals, from $5s$ through $11s$ and are shown in Fig. \ref{fig2}. For comparison, we have also shown the DF contributions from these individual orbitals. The correlation effects for the $6s$ orbital are much larger than its DF contribution itself. The core electrons of $d$ symmetry also influence the RCC amplitudes indirectly.  High accuracy calculations should therefore employ an all-order method such as the RCC theory to account for the strong correlation effects in a comprehensive manner.

We have also investigated the role of different doubly excited states that 
contribute through ${D}\, S_{2v}^{(1)}$. These contributions are given in
Table \ref{tab2}. As seen from the table, the largest contribution
comes when the $6s$ orbital is excited to the 
valence orbital, $6p_{1/2}$ followed by the latter's fine structure partner, the $6p_{3/2}$ orbital. There are also, however, some non-negligible contributions coming from the $5d$ orbitals when they get excited to the virtual $p_{3/2}$ and $f_{5/2,7/2}$ orbitals.
\begin{table}[h]
 \caption{Contributions from selected doubly excited states through ${D}\, S_{2v}^{(1)}$ term.}\label{tab2}
\begin{center}
\begin{tabular}{lccc|ccc}
\hline \hline
core($a$) & virtual($p$) & Results & & core($a$) & virtual($p$) & Results \\
\hline
6s & 6p$_{1/2}$ & 170.88 & & 5d$_{3/2}$ & 8f$_{5/2}$ & 0.92\\ 
6s & 7p$_{1/2}$ & 3.61 & & 5d$_{3/2}$ & 9f$_{5/2}$ & 2.06 \\ 
6s & 8p$_{1/2}$ & 2.01 & & 5d$_{5/2}$ & 6p$_{3/2}$ & 5.37 \\ 
6s & 9p$_{1/2}$ & 4.50 & & 5d$_{5/2}$ & 7p$_{3/2}$ & 0.76 \\ 
6s & 6p$_{3/2}$ & 31.42 & & 5d$_{5/2}$ & 8p$_{3/2}$ & 0.59 \\ 
6s & 7p$_{3/2}$ & 3.00 & & 5d$_{5/2}$ & 9p$_{3/2}$ & 1.46\\ 
6s & 8p$_{3/2}$ & 1.95 & & 5d$_{5/2}$ & 8f$_{7/2}$ & 2.34 \\ 
6s & 9p$_{3/2}$ & 5.26 & & 5d$_{5/2}$ & 9f$_{7/2}$ & 5.36 \\ 
\hline \hline 
\end{tabular}
\end{center}
 \end{table}

The EDM enhancement factor result for Tl from different calculations are compared in Table \ref{ef-tl-comp}. It is quite apparent that the published results lie in a rather large range. Although, the overall trends of the majority of the correlation terms in \cite{liu92} appear similar to ours, there is a significant difference in the magnitude of many of those correlation contributions. This can be due to the several approximations considered in their calculation, some of which are the following:
(i) they have considered only the one-body form for the unperturbed atomic Hamiltonian and as well as for the EDM Hamiltonian thereby neglecting the important contributions partly from the DF potential and 
largely from the two electron Coulomb interaction. (ii) they have considered 
only the linear terms and a few selected non-linear terms in their calculations. (iii) they have included selective triple excitations approximately in the unperturbed singles amplitude equations only (i.e., in $T_1^{(0)}$ and $S_{1v}^{(0)}$ equations), where as, they are completely 
ignored in the unperturbed doubles equations. 
(iv) they have frozen the inner core up to the $4 s$ orbital for the calculation of unperturbed amplitudes, where as, for solving the perturbed doubles equations they further freeze $4 s$, $4 p$
and $4 d$ orbitals. In contrast to the above set backs, we consider all the
non-linear terms arising from the single and double excitations. We also consider the leading triple excitations in both the
unperturbed singles and doubles cluster equations. We solve both the unperturbed and perturbed, closed- and open-shell equations, self-consistently. We have performed the all-electron relativistic CCSD(T) calculation and the total active orbitals considered here, are: $14s$, $13p_{1/2,3/2}$, $13d_{3/2,5/3}$, $9f_{5/2,7/2}$
and $8g_{7/2,9/2}$.

The comparison between the results of Dzuba and Flambaum based on a combined CI+MBPT approach \cite{dzuba2009} and our all-order CCSD(T) is not straight forward. They have considered the three outer shell electrons ($6s^2\,6p_{1/2}$) as valence and the rest as core. The core, virtual and valence orbitals are generated in a $V^{N-3}$ closed-shell potential in contrast to the $V^{N-1}$ potential used in our calculation. The valence-valence correlations are evaluated by CI, while the valence-core and the core-core correlations by MBPT. In view of the strong correlation effects in Tl, it is important to treat them by an all-order correlation method instead of dealing with it as a three valence system within the framework of an hybrid CI and finite order MBPT approach. It appears from the previous work of Dzuba and Flambaum that the ${\mathcal P}$ \& ${\mathcal T}$ violating Hamiltonian used in \cite{dzuba2009}, considers only the internal electric field due to the nucleus and not the electrons. The major drawback of the latter work is that all the correlations obtained by MBPT (except for s-electrons in the one-body correlation operator $\Sigma_1$ \cite{dzuba2009}) are considered only up to second order although they have emphasized the importance of valence-core correlations in their paper. In contrast, we have considered all these effects to all-orders in the residual Coulomb interaction in the framework of the full fledged CCSD(T) theory. The CI+MBPT result of \cite{dzuba2009} compares well with that of  \cite{liu92}. The latter as mentioned earlier is based on a linearized CCSD(T) approach with several approximations. However, the agreement between these two results is fortuitous.

\begin{table}[h]
\begin{center}
 \caption{Comparison of the ground state EDM EF (${\mathcal R}$) of Tl from different calculations.}\label{ef-tl-comp}
\begin{tabular}{lll}
\hline \hline 
  ${\mathcal R}$ & Method & Reference \\ \hline 
$-466(10)$   & CCSD(T) & This Work \\
$-582(20)$ & CI+MBPT & Dzuba 2009 \cite{dzuba2009}\\
$-585(30-60)$   & LCCSD(T) & Liu 1992 \cite{liu92} \\
$-179$   & MBPT(2) & Hartley 1990\cite{hartley90}\\
$-301$   & MBPT(2) & Kraftmakher 1988\cite{kraftmakher88} \\
$-502, -562, -607, -1041$  & MBPT(1) & Johnson 1986 \cite{johnson86} \\
$-500$   & semi-emp. & Flambaum 1976 \cite{flambaum76} \\
$-716$   & semi-emp. & Sandars 1975 \cite{sandars75}\\ \hline \hline
\end{tabular}
\end{center}
 \end{table}
We have also compared our result with a few \emph{ab initio}  
finite-order MBPT results \cite{johnson86,hartley90,kraftmakher88} and semi-empirical estimations \cite{flambaum76,sandars75} in
Table \ref{ef-tl-comp}. Apparently, the MBPT calculations \cite{hartley90, kraftmakher88} have under estimated the results, where as, the first-order MBPT calculation with semi-empirical potentials \cite{johnson86} and the semi-empirical calculations \cite{flambaum76,sandars75} have over estimated the EDM EF of Tl. 

The results of a few relevant physical quantities such as the electric dipole (E1) matrix element for $7S \rightarrow 6P_{1/2}$ transition and the magnetic dipole hyperfine structure constants of the ground state $6P_{1/2}$ and the lowest singly excited state $7S$ have been presented in Table \ref{e1-hfs}. It is clear that, our results are in better 
agreement with the experiments than those reported in \cite{dzuba2009}.
\begin{table}[h]
\begin{center}
 \caption{Comparison of the E1 matrix element and hyperfine structure constants of low-lying states in Tl.}\label{e1-hfs}
\begin{tabular}{cccc}
\hline \hline
           & This Work & Dzuba \& Flambaum \cite{dzuba2009} & Expt. \\
\hline
Transition & \multicolumn{3}{c}{E1 tr. amp.} \\ \cline{2-4}
$7\,S \rightarrow 6\,P_{1/2}$ & 1.82 & 1.73  & 1.81(2) \cite{hsieh1972}\\
State             & \multicolumn{3}{c}{Magnetic dipole hfs constant} \\ \cline{2-4} 
$6\,P_{1/2}$        & 21053 & 21067  & 21311 \cite{grexa1988}\\
$7\,S$        & 11894 & 11417  & 12297 \cite{herman1993}\\
\hline \hline 
\end{tabular}
\end{center}
 \end{table}

The error estimates quoted in \cite{dzuba2009} for Tl 
EDM EF does not seem to be convincing for the reason that 
they have treated the valence-core correlations for the most part only up to second order in perturbation theory. Clearly an all-order correlation calculation is needed 
for a reliable error estimation. Although, in the CCSD(T) approach the effect of the dominant triple excitations are considered while evaluating the CCSD amplitudes, their direct contributions are ignored while computing Eq. (\ref{enfact}). To estimate the maximum possible error on the EF result, we evaluate the contribution coming from the triple excitations, such as, $S_{vab}^{pqr,(0)}$ and $S_{vab}^{pqr,(1)}$ where the latter is defined as (for details, see \cite{sahoo2}),
\begin{eqnarray}
S_{vab}^{pqr,(1)} &=&  \frac{\widehat{H_0\, T_2^{(1)}} + \widehat{H_0\, S_{2v}^{(1)}}}{\epsilon_v + \epsilon_a + \epsilon_b - \epsilon_p - \epsilon_q - \epsilon_r}.
\end{eqnarray}
Considering the error due to the basis set incompleteness and the error due to the neglected triples, we obtain the final result for the EDM EF of Tl to be $-466(10)$. By combining the EDM experimental result for Tl, $-(4.0 \pm 4.3)\times 10^{-25}\, {\mathrm e\, cm}$, given by 
Regan et al. \cite{regan02} and our accurate theoretical EDM EF, we obtain a new limit for the electron EDM to be  $(8.7 \pm 9.3)\times 10^{-28}\, {\mathrm e\, cm}$ which translates into an upper limit, $d_e < 2.0 \times 10^{-27}\,{\mathrm e \, cm}$ at $90\%$ confidence level. In conclusion, we have obtained the most accurate limit for the electron EDM to date by improving the value of the EDM EF for the ground state of atomic Tl by treating the unusually strong electron correlation effects to all-order using the relativistic CCSD(T) method.

We thank the staff of CDAC, Bangalore for the useful discussions on some of the computational aspects related to this work and also for allowing us to use their facility. HSN thanks Dr. B. C. Regan for helpful discussions and acknowledges the kind hospitality of Dr. Mihaly Kallay at Budapest University of Technology and Economics.

\end{document}